\documentclass{intech08}

\booktitle{Will-be-set-by-IN-TECH}%

\chaptertitle{Stellar Nucleosynthesis Nuclear Data Mining} % You know your chapter title?

\authors{Boris Pritychenko}
\affiliation{National Nuclear Data Center, Brookhaven National Laboratory, Upton, NY 11973-5000}
\country{USA}

\begin{document}

\maketitle

\section{Introduction }
In the past 100 years, astronomy, astrophysics and cosmology have evolved from the observational and theoretical fields into more experimental science, when many stellar and planetary processes are recreated in physics laboratories and extensively studied \citep{Boyd09}. Many astrophysical phenomena have been explained using our understanding of nuclear physics processes, and the whole concept of stellar nucleosynthesis has been introduced. The importance of nuclear reactions as a source of stellar energy was recognized by Arthur Stanley Eddington as early as 1920 \citep{Eddington20}. Later, nuclear mechanisms by which hydrogen is fused into helium were proposed by Hans Bethe \citep{Bethe39}. However, neither of these contributions explained the origin of  elements heavier than helium.

Further developments helped to identify the Big Bang, stellar and explosive nucleosynthesis  processes that are responsible for the currently-observed variety of elements and isotopes \citep{Hoyle46,Merrill52,Burbidge57,Cameron57}. Today, nuclear physics is successfully applied to explain the variety of elements and isotope abundances observed in stellar surfaces, the solar system and cosmic rays via network calculations and comparison with observed values.

A comprehensive analysis of stellar energy production, metallicity and isotope abundances   indicates the crucial role of proton-, neutron- and light ion-induced nuclear reactions and $\alpha$-, $\beta$-decay rates. These subatomic processes govern the observables and predict the star life cycle. Calculations of the transition rates between isotopes in a network strongly rely on theoretical and experimental cross section and decay rate values at stellar temperatures. Consequently, the general availability of nuclear data is of paramount importance in stellar nucleosynthesis research.

This chapter will provide a review of theoretical and experimental nuclear reaction and structure data for stellar and explosive nucleosynthesis and modern computation tools and methods. Examples of evaluated and compiled nuclear physics data will be given. Major nuclear databases and their input for nucleosynthesis calculations will be discussed.
%The body of manuscript begins with Introduction (1st page) and ends with the last reference\footnote{This is footnote} at the bottom of last page of manuscript. The size of manuscript must be exactly 8 or 10 or 12 or 14 or 16 or 18 or 20 or 22 or 24 or 26 full pages. (Book Antiqua, 9pt, normal)

\section{Nucleosynthesis and its data needs}
Nucleosynthesis is an important nuclear astrophysics phenomenon that is responsible for presently observed chemical elements and isotope abundances. It started in the early Universe and presently proceeds in the stars. The Big Bang nucleosynthesis is responsible for a relatively high abundance  of the lightest primordial elements in the Universe from $^{1}$H to $^{7}$Li, and it precedes  stars formation and stellar nucleosynthesis. The general consistency between theoretically predicted and observed lightest elements abundances serves as a strong evidence for the Big Bang theory \citep{Kolb88}.

\begin{figure}[htb]	% h-here, t-top, b-bottom
\centering
\includegraphics[width=12cm]{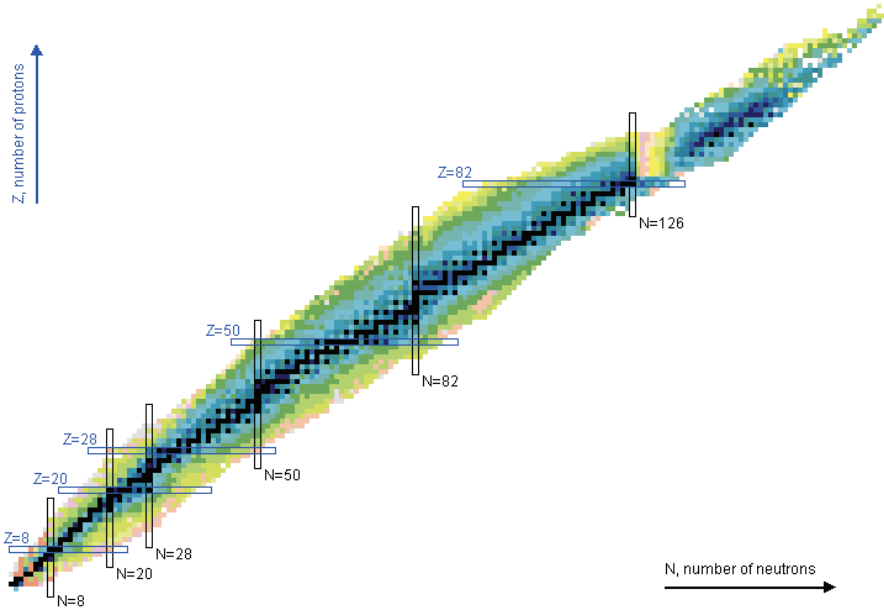} %	** if .eps don't need extension
\caption{The chart of nuclides. Stable and long-lived ($>$10$^{15}$ s) nuclides  are shown in black. Courtesy of NuDat Web application ({\it http://www.nndc.bnl.gov/nudat}).}\label{chart}
\end{figure}

\begin{figure}[htb]	% h-here, t-top, b-bottom
\centering
\includegraphics[width=12cm]{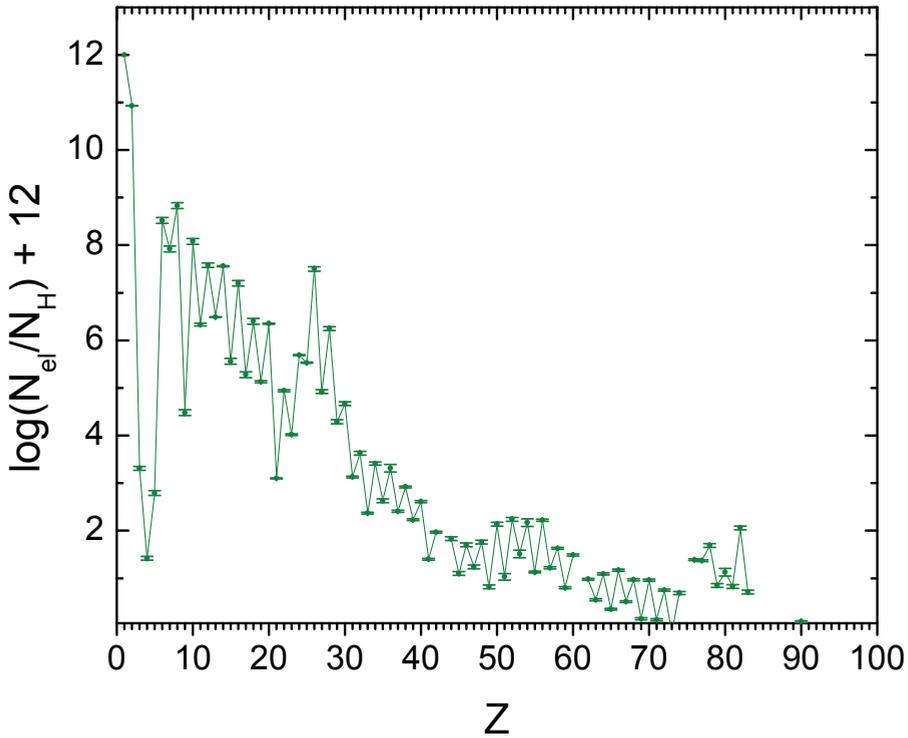} %	** if .eps don't need extension
\caption{Solar system elemental abundances; data are taken from \citep{Grevesse98}.}\label{abundances}
\end{figure}

The currently-known variety of nuclei and element abundances is shown in Fig. \ref{chart} and Fig. \ref{abundances}. These Figures indicate a large variety of isotopes in Nature \citep{Anders89} and the strong need for additional nucleosynthesis mechanisms beyond the Big Bang theory.

These additional mechanisms have  been pioneered by Eddington  \citep{Eddington20} via introduction of a revolutionary concept of element production in the stars. Present nucleosynthesis models  explain medium and heavy element abundances using the stellar nucleosynthesis that consists of  burning (explosive)  stages of stellar evolution, photo disintegration, and neutron and proton capture processes. The model predictions can be verified through the star metallicity studies and comparison of calculated isotopic/elemental abundances with the observed values, as shown in Fig. \ref{abundances} .

Nowadays, there are many well-established theoretical models of stellar nucleosynthesis  \citep{Burbidge57,Boyd09}; however, they still cannot reproduce the observed abundances due to many parameter uncertainties.  The flow of the nuclear physics processes in the network calculations is defined by the nuclear masses, reaction, and decay rates, and strongly correlated with the stellar temperature and density. These calculations depend heavily on our understanding of nuclear physics  processes in stars, and the availability of high quality nuclear  data.

The common sources of stellar nucleosynthesis data include KADONIS, NACRE and REACLIB dedicated nuclear astrophysics libraries \citep{Dillmann06,Angulo99,Cyburt10} and general nuclear science and industry databases. Dedicated libraries are optimized for nuclear astrophysics applications, and contain pre-selected data that are often limited to the original scope. In many cases, these sources reflect the present state of nuclear physics, when experimental data are not always available or limited to a single measurement. 	Such limitation highlights the importance of theoretical calculations that strongly dependent on nuclear models. Another problem arises from the fact that nucleosynthesis processes are strongly affected by the astrophysical site conditions.

To broaden the scope of the traditional nuclear astrophysics calculations,  we will investigate applicability of nuclear physics databases for stellar nucleosynthesis data mining. These databases were developed for nuclear science, energy production and national security applications and will provide  complementary astrophysics model-independent results.
%Manuscript must contain clear answers to following questions: What is the problem / What has been done by other researchers and where you can contribute / What have you done / Which method or tools you used / What are your results / What is new and good, what is not good / Future research

\section{Stellar nucleosynthesis}
This section will briefly consider the light elements and concentrate on the active research subject of production of medium and heavy elements beyond iron via slow and rapid neutron  capture and associated data needs.  Finally, photo disintegration and proton capture nucleosynthesis processes will be reviewed.

\subsection{Burning phases of stellar evolution}
\begin{figure}[htb]	% h-here, t-top, b-bottom
\centering
\includegraphics[width=13cm]{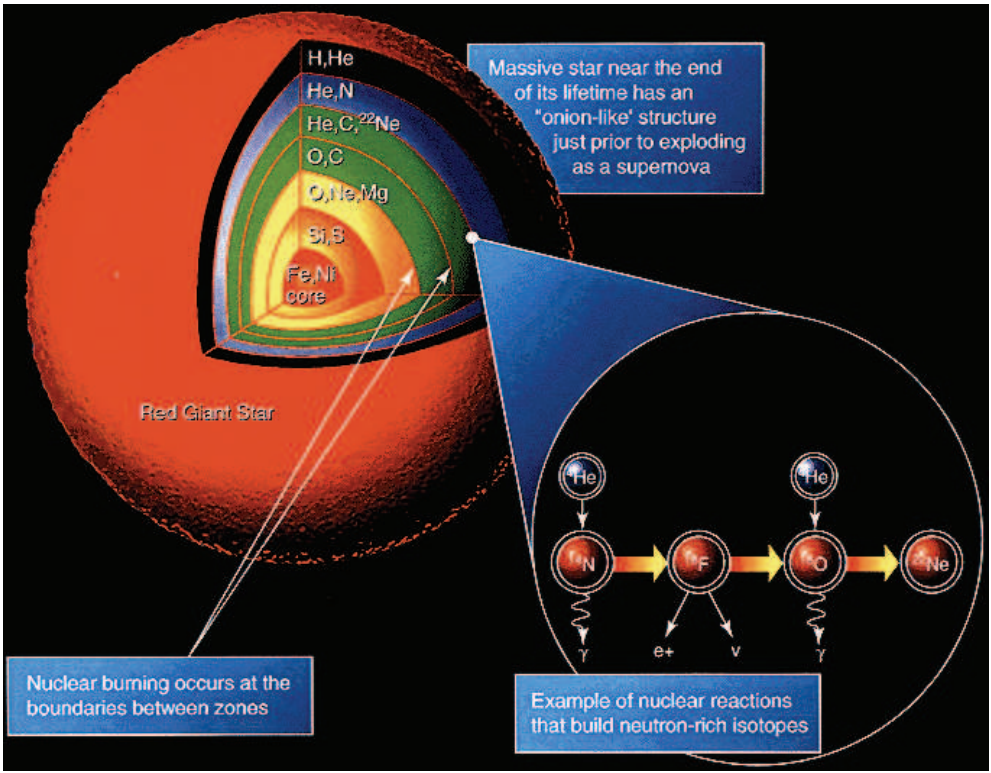} %	** if .eps don't need extension
\caption{Cross section of a Red Giant showing nucleosynthesis and elements formed. Courtesy of  Wikipedia ({\it http://en.wikipedia.org/wiki/Stellar$\_$nucleosynthesis}).}\label{red}
\end{figure}

Fusion reactions are responsible for burning phases of stellar evolution. These reactions produce light, tightly-bound nuclei and release energy. The process of new element creation proceeds before nuclear binding energy reaches maximum value in the Fe-Ni region. Four important cases will be reviewed: pure hydrogen burning, triple alpha process, CNO cycle, and stellar burning. These processes take place in stars with a mass similar to our Sun, as shown in Fig. \ref{red}.  The data needs for these processes are addressed in the IAEA FENDL  \citep{Aldama2004} and EXFOR  (Experimental Nuclear Reaction Data) \citep{NRDC11} databases.

\subsubsection{Pure hydrogen burning}
Hydrogen is the most abundant element in the Universe. The proton-proton chain dominates stellar nucleosynthesis in stars comparable to our Sun
\begin{equation}
p + p \rightarrow d + \beta^{+}  + \nu + 0.42 MeV
\end{equation}
Further analysis of the pp-process \citep{Burbidge57} indicates extremely low cross sections for E$<$1 MeV nuclear projectiles and explains the necessity of large target mass and density for sustainable nuclear fusion reaction
\begin{equation}
\sigma \sim 3 E^{4.5}  \times 10^{-24} b
\end{equation}
Next, the deuterium produced in the first stage can fuse with another hydrogen
\begin{equation}
d  +  p  	\rightarrow  ^{3}He  + \gamma  + 5.49 MeV	
\end{equation}
Finally, $^{4}$He will be produced in the pp I and pp II branches.

\subsubsection{Triple alpha process}
The hydrogen burning in stars leads to the production of the helium core at a star's center. Further helium burning goes through the $3\alpha$ process
\begin{eqnarray}
\alpha + \alpha &  \Leftrightarrow  &  ^{8}Be  -93.7  keV \nonumber  \\
^{8}Be + \alpha & \Leftrightarrow & ^{12}C^{*} + 7.367 MeV
\end{eqnarray}

\subsubsection{CNO cycle}
Carbon-Nitrogen-Oxygen cycle leads to the production of elements heavier than carbon and consists of major CNO-I (regeneration of carbon and alpha particle)
\begin{equation}
^{12}C \rightarrow ^{13}N \rightarrow ^{13}C \rightarrow ^{14}N \rightarrow ^{15}O \rightarrow ^{15}N \rightarrow ^{12}C
\end{equation}
and minor CNO-II branches (production of $^{16}$O and proton):
\begin{equation}
^{15}N  \rightarrow ^{16}O \rightarrow ^{17}F \rightarrow ^{17}O  \rightarrow ^{14}N \rightarrow ^{15}O \rightarrow ^{15}N
\end{equation}

\subsubsection{Advanced stages of stellar burning}
Further burning includes carbon, neon, oxygen and silicon burning. Here, we have high-temperature and density burning stages when photonuclear processes produce additional $\alpha$-particles and  create a complex network of reactions with light particles.  These reactions produce fusion nuclei up to the Fe-Ni peak. The Fe-Ni region nuclei are the most tightly bound in Nature and fusion reactions stop. Further element production proceeds via neutron and proton captures and photo disintegration.

\subsection{s-process}
The slow-neutron capture (s-process) is responsible for creation of $\sim$50 $\%$ of the elements beyond iron. In this region, neutron capture  becomes dominant because of the increasing Coulomb barrier and decreasing binding energies. This s-process takes place in the Red Giants and AGB stars, where neutron temperature ($kT$) varies from  8 to 90 keV. A steady supply of neutrons is available due to   H-burning and He-flash reactions
\begin{eqnarray}
^{13}C(\alpha,n)^{16}O  \nonumber  \\
^{22}Ne(\alpha,n)^{25}Mg
\end{eqnarray}
Fig. \ref{abundances} indicates a high abundance of  $^{56}$Fe nuclei due to termination of the explosive nucleosynthesis. It is natural to assume that iron  acts as a seed for the neutron capture reactions that eventually produce medium and heavy elements. The neutron capture time of s-process takes approximately one year. Consequently, the process path lies along the nuclear valley of stability up to the last long-lived  nucleus of $^{209}$Bi.

In the giant stars, neutron reaction rates define the elemental abundances, and in some cases branching points (created by the competition between neutron capture and $\beta$-decay) strongly affect the heavy isotope production rates.  Therefore, special attention has to be paid to the branching points at $^{79}$Se,$^{134}$Cs,$^{147}$Pm, $^{151}$Sm, $^{154}$Eu, $^{170}$Yb and $^{185}$W, and the neutron poison (absorption) $^{16,18}$O,$^{22}$Ne(n,$\gamma$) reactions. The s-process astrophysical site  conditions imply the following data needs
\begin{itemize}
\item Reaction rates
\begin{itemize}
\item neutron-induced
\item charged particle
\end{itemize}
\item Half-lives
\end{itemize}
Convincing proof of s-process existence and its role in Nature could come from the calculation of isotopic abundances and comparison with observed values \citep{Anders89,Grevesse98}. Present-day s-process nucleosynthesis calculations often are based on the dedicated nuclear astrophysics data tables, such as works of  \citep{Bao00}, and \citep{Rauscher00}. These data tables contain quality information on Maxwellian-averaged cross sections ($\langle \sigma^{Maxw}_{\gamma} (kT) \rangle$) and astrophysical reaction rates ($R(T_9)$). However, it is essential to produce complementary neutron-induced reaction data sets for an independent verification and to expand the boundaries of the existing data tables.

\begin{figure}[htb]	% h-here, t-top, b-bottom
\centering
\includegraphics[width=13cm]{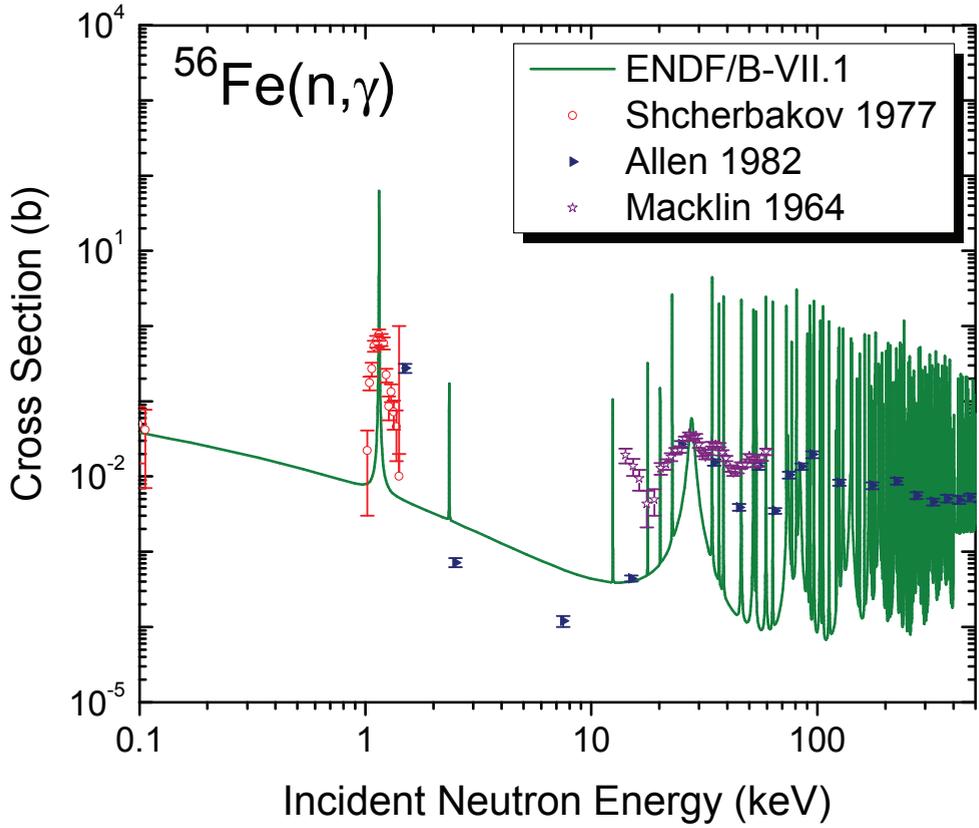} %	** if .eps don't need extension
\caption[ENDF/B-VII.1 and EXFOR libraries  $^{56}$Fe(n,$\gamma$) cross sections for astrophysical range of energies.]{ENDF/B-VII.1 and EXFOR libraries  $^{56}$Fe(n,$\gamma$) cross sections \citep{Chadwick06,NRDC11} for astrophysical range of energies.}\label{56FeCS}
\end{figure}

Recent releases of ENDF/B-VII evaluated nuclear reaction libraries \citep{Chadwick06} and publication of the Atlas of Neutron Resonances reference book \citep{Mughabghab06} created a unique opportunity of applying these data for non-traditional applications, such as s-process nucleosynthesis \citep{Pritychenko10}. Many neutron cross sections for astrophysical range of energies, including   $^{56}$Fe(n,$\gamma$)  reaction as shown in Fig. \ref{56FeCS}, are available in the ENDF (Evaluated Nuclear Data File) and EXFOR libraries. The feasibility study of the evaluated nuclear data for s-process nucleosynthesis will be presented below.

\subsubsection{Calculation of Maxwellian-averaged cross sections and uncertainties}
The Maxwellian-averaged cross section  can be expressed as
\begin{equation}
\label{myeq.max3}
\langle \sigma^{Maxw}(kT)  \rangle = \frac{2}{\sqrt{\pi}} \frac{(m_2/(m_1 + m_2))^{2}}{(kT)^{2}}  \int_{0}^{\infty} \sigma(E^{L}_{n})E^{L}_{n} e^{- \frac{E^{L}_{n} m_2}{kT(m_1 + m_2)}} dE^{L}_{n},
\end{equation}
where {\it k} and {\it T} are the Boltzmann constant and temperature of the system, respectively,  and $E$ is an energy of
relative motion of the neutron with respect to the target. Here,  $E^{L}_{n}$ is a neutron energy in the laboratory system
and $m_{1}$ and $m_{2}$ are masses of a neutron and target nucleus, respectively.

The astrophysical reaction rate for network calculations is defined as
\begin{equation}
\label{rrate}
R(T_9) = N_{A}\langle\sigma v\rangle = 10^{-24}  \sqrt(2 kT/\mu)  N_{A}  \sigma^{Maxw}(kT),
\end{equation}
where $T_9$ is temperature expressed in billions of Kelvin, $N_A$ is an Avogadro number. $T_9$ is related to the   $kT$ in  MeV units as follows
\begin{equation}
\label{t9}
11.6045 \times T_9 = kT
\end{equation}
It is commonly known that  for the equilibrium {\it s}-process-only nuclei product  of $\langle \sigma^{Maxw}_{\gamma} (kT) \rangle$ and solar-system abundances ($N_{(A)}$)  is preserved \citep{Rolfs88}
\begin{equation}
\label{myeq.s0}
\sigma_{A}N_{(A)}= \sigma_{A-1}N_{(A-1)} = constant
\end{equation}
The stellar equilibrium conditions provide an important test for the s-process nucleosynthesis in mass regions between neutron magic numbers N=50,82,126 \citep{Arlandini99}. To investigate this phenomenon,  we will consider  ENDF libraries. These data were never adjusted for nuclear  astrophysics models and  are essentially model-independent.

\begin{figure}[htb]	% h-here, t-top, b-bottom
\centering
\includegraphics[width=13cm]{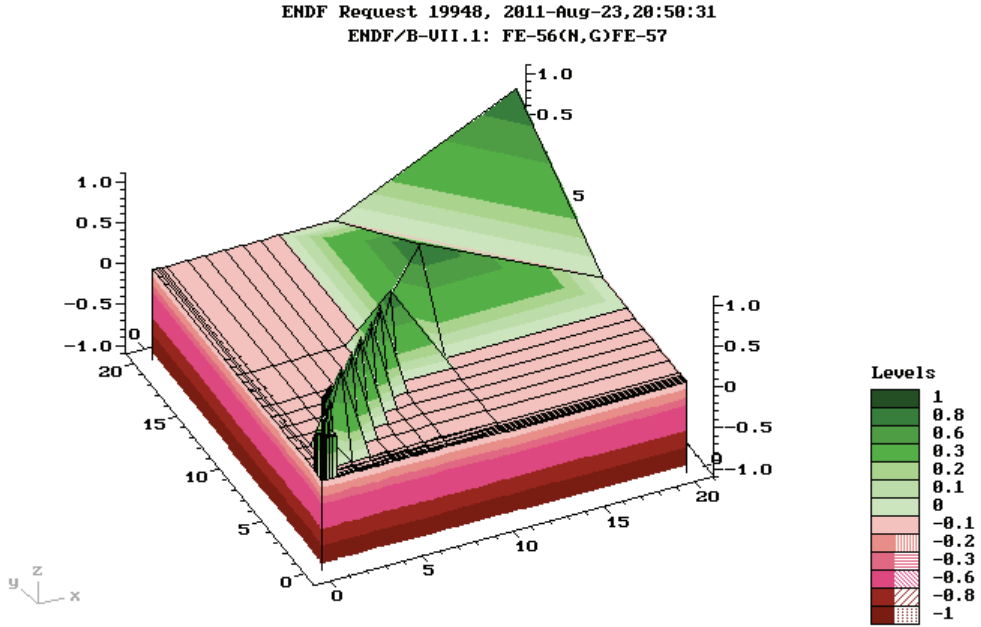} %	** if .eps don't need extension
\caption[ENDF/B-VII.1 library  $^{56}$Fe(n,$\gamma$) cross section covariances.]{ENDF/B-VII.1 library   $^{56}$Fe(n,$\gamma$) cross section covariances \citep{Zerkin05}.}\label{56FeCov}
\end{figure}

ENDF library is a core nuclear reaction database containing evaluated (recommended) cross sections, spectra, angular distributions, fission product yields, thermal neutron scattering, photo-atomic and other data, with emphasis on neutron-induced reactions.  ENDF library evaluations cover all neutron reaction channels within 10$^{-5}$ eV - 20 MeV energy range.  In many cases, evaluations contain information on neutron cross section covariances.  An example of ENDF/B-VII.1library  $^{56}$Fe(n,$\gamma$) neutron cross section covariances (uncertainties) is  shown in  Fig. \ref{56FeCov}.

The evaluated neutron libraries are based on theoretical calculations using EMPIRE, TALYS and Atlas collection of nuclear reaction model codes \citep{Herman07,Koning08,Mughabghab06} that are often adjusted to fit experimental data \citep{NRDC11}. The model codes are essential for neutron cross section calculations of short-lived radioactive nuclei where  experimental data are not available. The ENDF data files are publicly available from the NNDC (National Nuclear Data Center) Sigma Web Interface: {\it http://www.nndc.bnl.gov/sigma} \citep{Pritychenko08}.

\begin{figure}[htb]	% h-here, t-top, b-bottom
\centering
\includegraphics[width=12.5cm]{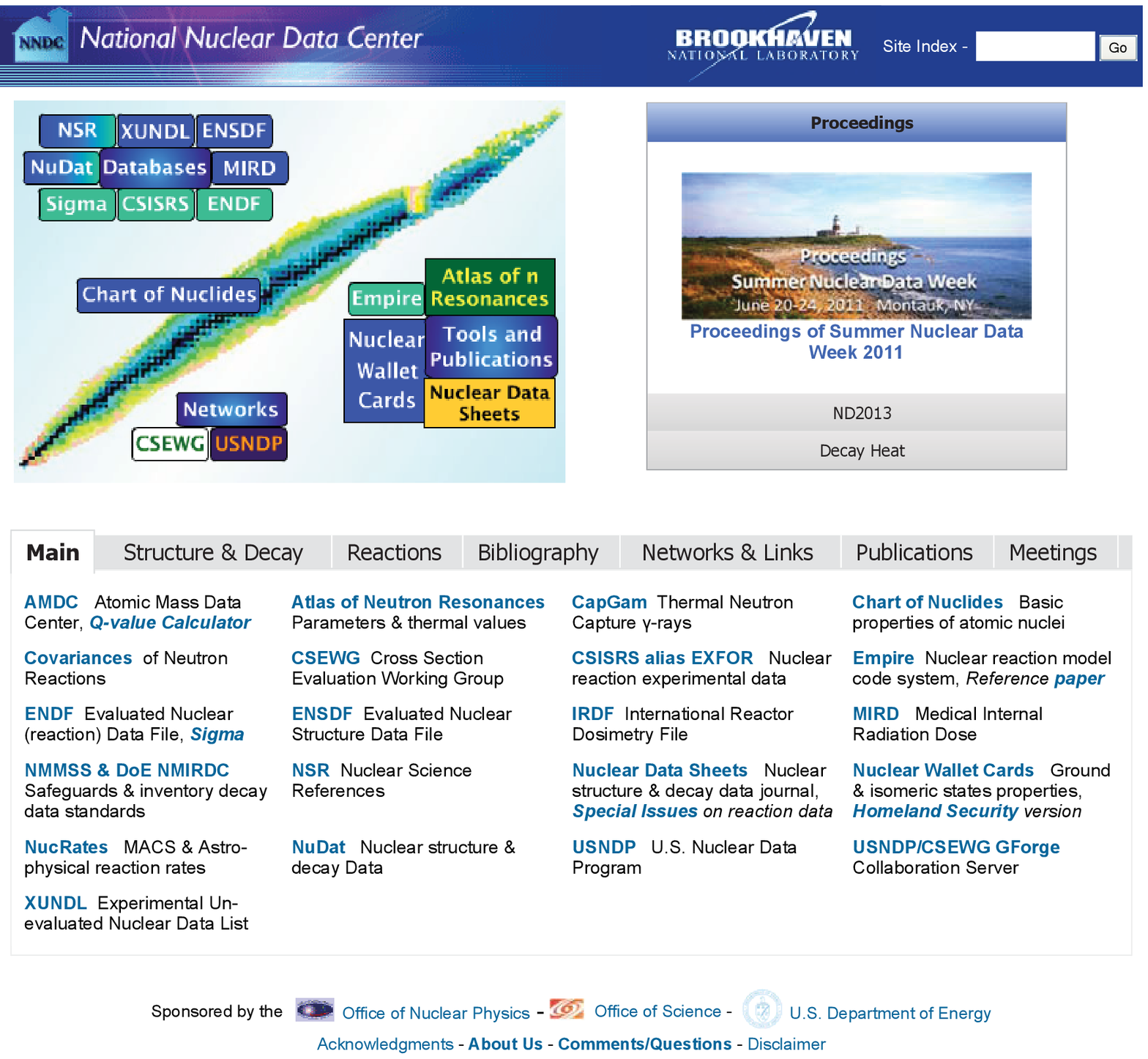} %	** if .eps don't need extension
\caption{The NNDC website  ({\it http://www.nndc.bnl.gov}) provides access to nuclear databases and Maxwellian-averaged cross sections and astrophysical reaction rates online calculations.}\label{nucrates}
\end{figure}

Previously, Maxwellian-averaged cross sections and astrophysical reaction rates were produced using the Simpson method for the linearized ENDF cross sections  \citep{Pritychenko10} . A similar effort was  completed at the Japanese Atomic Energy Agency  \citep{Nakagawa05}.  Fig.  \ref{nucrates} shows the NNDC  website \citep{Pritychenko06} that provides access to nuclear databases and  the NucRates Web application ({\it http://www.nndc.bnl.gov/astro}).  The NucRates application was designed for  online calculations of $\langle \sigma^{Maxw}(kT)  \rangle$  and $R(T_9)$ using all major evaluated nuclear reaction libraries, astrophysical neutron-induced reactions and {\it kT} values ranging from 10$^{-5}$ eV to 20 MeV.

The Simpson method allowed quick calculation of integral values; however, the degree of precision was within $\sim$1$\%$.  Precision can be improved with the linearized ENDF files because the cross section value is linearly-dependent on energy within a particular bin \citep{BPritychenko10}
\begin{equation}
\label{myeq.int1}
\sigma (E) = \sigma (E_1) + (E-E_1)\frac{\sigma (E_2) - \sigma (E_1)}{E_2 - E_1},
\end{equation}
where $\sigma (E_1),  E_1$ and  $\sigma (E_2),  E_2$ are cross section and energy values for the corresponding energy bin. The last equation is a good approximation of neutron cross section values for a sufficiently dense energy grid. This allowed deduction of  $\langle \sigma^{Maxw}_{\gamma} (kT) \rangle$ definite integrals for separate energy bins  using Doppler-broadened  cross sections and the Wolfram Mathematica online integrator \citep{Wolf11}. Further, summing integrals for all energy bins will produce a precise ENDF value for the Maxwellian-averaged cross section.

\begin{figure}[htb]	% h-here, t-top, b-bottom
\centering
\includegraphics[width=10cm]{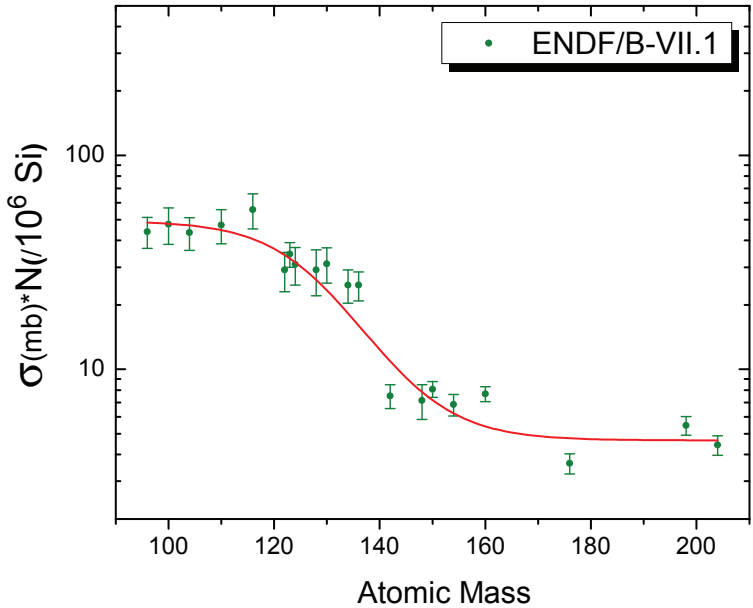} %	** if .eps don't need extension
\caption[ENDF/B-VII.1 library product of neutron-capture cross section (at 30 keV in mb) times solar system abundances (relative to Si = 10$^6$) as a function of atomic mass for nuclei produced only in the {\it s}-process.]{ENDF/B-VII.1 library product of neutron-capture cross section (at 30 keV in mb) times solar system abundances (relative to Si = 10$^6$) as a function of atomic mass for nuclei produced only in the {\it s}-process.}\label{sigma}
\end{figure}

The product values of the ENDF/B-VII.1 $\langle \sigma^{Maxw}_{\gamma} (30 keV) \rangle$ times solar abundances \citep{Anders89} are plotted  in Fig.  \ref{sigma}.  They reveal that the ENDF/B-VII.1 library data closely replicate a two-plateau plot \citep{Rolfs88}.  The current result provides a powerful testimony for stellar nucleosynthesis.

\begin{figure}[htb]	% h-here, t-top, b-bottom
\centering
\includegraphics[width=13cm]{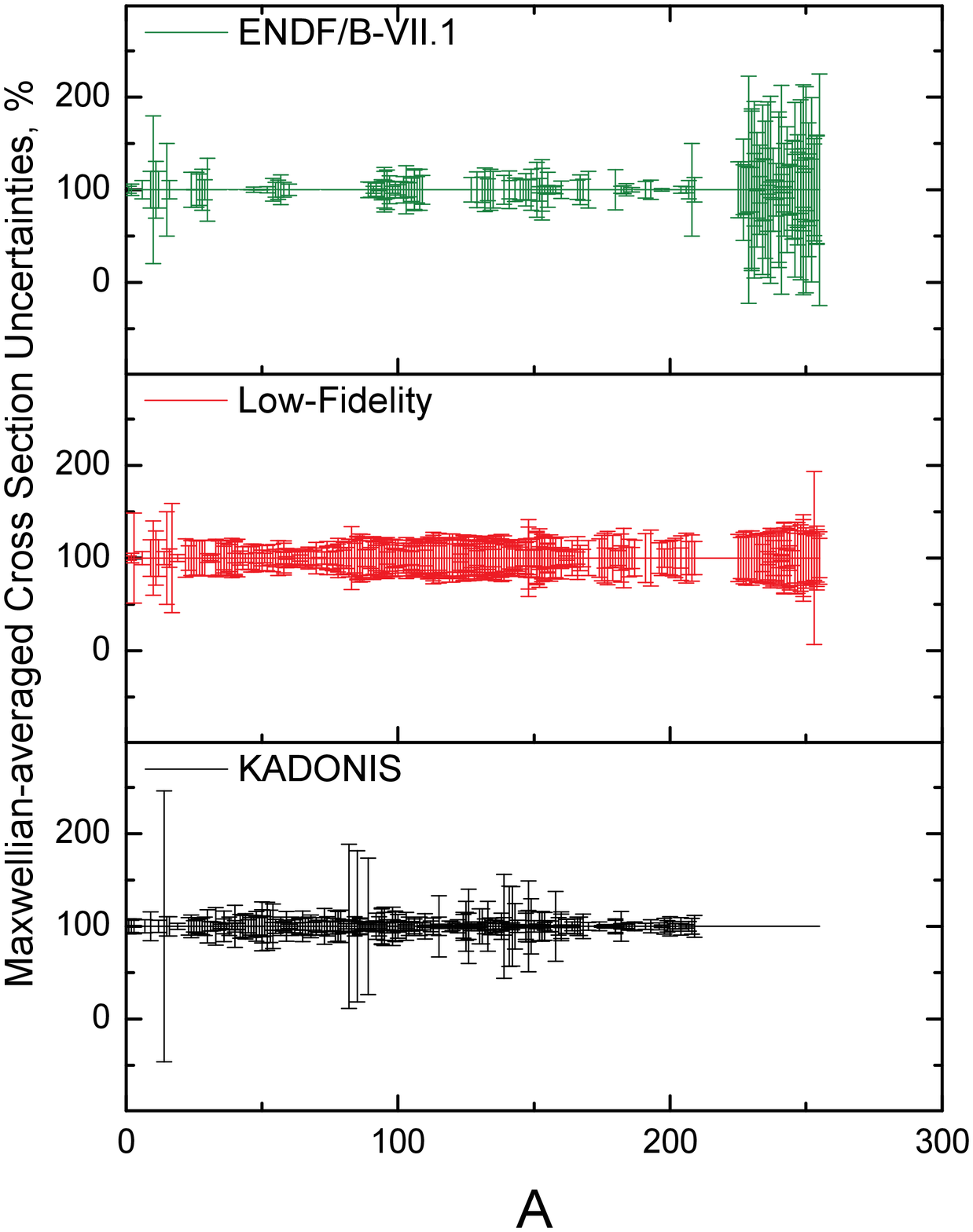} %	** if .eps don't need extension
\caption[Maxwellian-averaged Cross Sections Uncertainty for ENDF/B-VII.1, Low-Fidelity and KADONIS projects.]{Maxwellian-averaged cross section uncertainties for ENDF/B-VII.1, Low-Fidelity and KADONIS libraries \citep{Chadwick06,Little08,Dillmann06}.}\label{MaxUn}
\end{figure}

The predictive power of stellar nucleosynthesis calculations depends heavily on the neutron cross section values and their covariances. To understand the unique isotopic signatures from the presolar grains, $\sim$1\% cross section uncertainties are necessary \citep{FKaeppeler11}. Unfortunately, present uncertainties are often much higher, as shown in Fig. \ref{MaxUn}.

The situation gets even more complex after considering s-process branches where  $\beta$-decay and neutron capture rates are
\begin{equation}
\label{myeq.br}
\lambda_{n} \sim \lambda_{\beta}
\end{equation}
Here, the stellar thermal environment may affect  $\beta$-decay rates and change the process path. Branching is particularly important for unstable isotopes such as  $^{134}$Cs. This isotope can either decay to $^{134}$Ba, if neutron flux is low or capture neutron and produce $^{135}$Cs. The  $^{134}$Cs $\beta$-decay lifetime may vary from $\sim$1 y to 30 day over the temperature range of (100-300)$\times$10$^6$ K, which further complicates the calculations.

Finally, the  calculated values of Maxwellian-averaged cross sections at {\it kT}=30 keV, for selected s-process nuclei,     derived from the JENDL-4.0, ROSFOND 2010 and ENDF/B-VII.1  libraries \citep{Shibata01,Zabrodskaya07,Chadwick06} have been produced and  shown in Table \ref{Table1}.  The tabulated results are compared with the KADONIS values \citep{Dillmann06}.  Due to a limited number of ENDF covariance files, Low-Fidelity cross section covariances  \citep{Little08} were used to calculate  uncertainties for ENDF/B-VII.1 data.

The complete sets of ENDF s-process nucleosynthesis data sets are available for download from the NucRates Web application {\it http://www.nndc.bnl.gov/astro}. These complimentary data sets demonstrate a strong correlation between nuclear astrophysics and nuclear industry data needs,  the large nuclear astrophysics potential of ENDF libraries, and a perspective beneficial relationship between both fields.
\begin{table}
\centering
\begin{tabular}{|c|cccc|}
\hline
Isotope &  JENDL-4.0  &   ROSFOND 2010  &  ENDF/B-VII.1    &  KADONIS \\
\hline \hline
 42-Mo- 96 & 1.052E+2 &   1.035E+2 &  1.035E+2$\pm$1.700E+1   &  1.120E+2$\pm$8.000E+0 \\
44-Ru-100 &  2.065E+2 &   2.062E+2 &  2.035E+2$\pm$3.949E+1   &  2.060E+2$\pm$1.300E+1 \\
46-Pd-104 &  2.700E+2 &   2.809E+2 &  2.809E+2$\pm$4.923E+1   &  2.890E+2$\pm$2.900E+1 \\
48-Cd-110 &  2.260E+2 &   2.346E+2 &  2.349E+2$\pm$4.263E+1   &  2.370E+2$\pm$2.000E+0 \\
50-Sn-116 &  9.115E+1 &   1.002E+2 &  1.003E+2$\pm$1.875E+1   &  9.160E+1$\pm$6.000E-1 \\
52-Te-122 &  2.644E+2 &   2.639E+2 &  2.349E+2$\pm$4.882E+1   &  2.950E+2$\pm$3.000E+0 \\
52-Te-123 &  8.138E+2 &   8.128E+2 &  8.063E+2$\pm$1.063E+2   &  8.320E+2$\pm$8.000E+0 \\
52-Te-124 &  1.474E+2 &   1.473E+2 &  1.351E+2$\pm$2.697E+1   &  1.550E+2$\pm$2.000E+0 \\
54-Xe-128 &  2.582E+2 &   2.826E+2 &  2.826E+2$\pm$6.823E+1   &  2.625E+2$\pm$3.700E+0 \\
54-Xe-130 &  1.333E+2 &   1.518E+2 &  1.518E+2$\pm$2.835E+1   &  1.320E+2$\pm$2.100E+0 \\
56-Ba-134 &  2.301E+2 &   2.270E+2 &  2.270E+2$\pm$4.038E+1   &  1.760E+2$\pm$5.600E+0 \\
56-Ba-136 &  7.071E+1 &   7.001E+1 &  7.001E+1$\pm$1.087E+1   &  6.120E+1$\pm$2.000E+0 \\
60-Nd-142 &  3.557E+1 &   3.701E+1 &  3.343E+1$\pm$4.251E+1   &  3.500E+1$\pm$7.000E-1 \\
62-Sm-148 &  2.361E+2 &   2.444E+2 &  2.449E+2$\pm$4.507E+1   &  2.410E+2$\pm$2.000E+0 \\
62-Sm-150 &  4.217E+2 &   4.079E+2 &  4.227E+2$\pm$3.607E+2   &  4.220E+2$\pm$4.000E+0 \\
64-Gd-154 &  9.926E+2 &   1.010E+3 &  9.511E+2$\pm$1.096E+2   &  1.028E+3$\pm$1.200E+1 \\
66-Dy-160 &  8.702E+2 &   8.293E+2 &  8.328E+2$\pm$6.769E+1   &  8.900E+2$\pm$1.200E+1 \\
72-Hf-176 &  5.930E+2 &   4.529E+2 &  4.531E+2$\pm$4.896E+1   &  6.260E+2$\pm$1.100E+1 \\
80-Hg-198 &  1.612E+2 &   1.612E+2 &  1.613E+2$\pm$1.635E+1   &  1.730E+2$\pm$1.500E+1 \\
82-Pb-204 &  8.355E+1 &   7.242E+1 &  7.242E+1$\pm$7.624E+0   &  8.100E+1$\pm$2.300E+0 \\
\hline
\end{tabular}
\caption[Evaluated nuclear reaction libraries and KADONIS  libraries Maxwellian-averaged neutron capture cross sections.]{Evaluated nuclear reaction and KADONIS libraries  Maxwellian-averaged neutron capture cross sections in mb at {\it kT}=30 keV for  {\it s}-process nuclei.}\label{Table1}
\end{table}

\subsection{r-process}
The detailed analysis of stable and long-lived nuclei  indicates the large number of isotopes  that lie outside of the s-process path peaks near A=138 and 208.  In addition, the large gap between s-process nucleus $^{209}$Bi and  $^{232}$Th,$^{235,238}$U effectively terminates the $s$-process at $^{210}$Po.  Production of the actinide neutron-rich nuclei cannot be explained by the $s$-process nucleosynthesis and requires introduction of  rapid neutron capture or r-process. In this case, neutron capture timescale has to be less than typical $\beta$-decay lifetimes of $\sim$ms for neutron-rich nuclides. It implies neutron fluxes 10$^{10}$-10$^{11}$ higher than those of the s-process. Such conditions can be found in $\nu$-driven  core-collapse supernova and neutron stars. From here, one may conclude that r-process temperature depends on the site and may lie within a (0.5-10) $\times$10$^{9}$ K range. However, it is still not clear where r-process takes place and how it proceeds.

Among many unknowns of  the r-process  is process path.  The path is defined by the nuclear masses  and $\beta$-decay half-lives. The 2003 \& 2011 experimental Atomic Mass Evaluations    \citep{Audi03,Audi11} do not cover nuclei far from stability near the r-process expected path. To resolve this problem, theoretical mass calculations based on the  FRDM  and other models have been performed with 25$\%$ uncertainties \citep{Aprahamian11}. This calculation helped to identify the list of critical nuclei along the r-process path and demonstrated strong connections between nuclear astrophysics and nuclear structure.  Further progress will require mass and half-life measurements of unstable nuclei. The r-process data needs can be summarized as follows
\begin{itemize}
\item nuclear masses
\item $\beta$-decay half-lives
\item n-capture rates
\item neutrino interaction rates
\item fission probabilities
\item fission products distribution
\end{itemize}
Recent r-process estimates \citep{Cowan04}  demonstrate sharp abundance peaks for the A=130,195 (N=82,126) nuclei with large N/Z ratios and another broad peak at A=160.  Further analysis of  r-process abundances  shows \citep{Boyd08}
\begin{equation}
\label{myeq.r1}
dN_{Z,A} /dt  = \lambda_{Z-1}N_{Z-1}  - \lambda_{Z} N_{Z}
\end{equation}
For neutron closed shells, nuclei most likely will experience $\beta$-decay rather than absorb another neutron
\begin{equation}
\label{myeq.r2}
\lambda_{Z}N_{Z}  = N_{Z} / \tau_{Z}
\end{equation}
Thus near neutron closed shells, the relationship between abundances and $\beta$-decay lifetimes is
\begin{equation}
\label{myeq.r3}
dN_{Z,A} /dt  = N_{Z-1,A-1} / \tau_{Z-1} - N_{Z,A} /\tau_{Z}
\end{equation}
Finally, for equilibrium conditions one can deduce r-process analog of the equation \ref{myeq.s0}
\begin{equation}
\label{myeq.r4}
N_{Z-1} / \tau_{Z-1}  = N_{Z} /\tau_{Z}
\end{equation}
From the last formula, one can conclude that closed shell nuclei with the largest half-lives will have the largest abundances.  In order to obtain the complete picture, all half-lives and decay modes have to be determined. A list of properties relevant to r-process A$\sim$130 nuclei is shown in  Table \ref{Table2}. It includes $\beta$-decay half-lives and emission probabilities for delayed neutrons. The delayed neutrons provide an additional neutron source for the r-process and may shift the location of the abundance peak. The tabulated data were taken from the Evaluated Nuclear Structure Data File (ENSDF) database \citep{Burrows90} and the relation between nuclear lifetime and half-life is
\begin{equation}
\label{myeq.r5}
 \tau = T_{1/2} / 0.693
 \end{equation}

\begin{table}
\centering
\begin{tabular}{|c|ccc|}
\hline
Isotope &  T$_{1/2}$, msec  & $\beta^{-}$-decay, $\%$  &  $\beta$-n Emission, $\%$ \\
\hline \hline
$^{127}$Ag &109$\pm$25  & 100 & ? \\
$^{128}$Ag & 58$\pm$5  & 100 & ? \\
$^{129}$Ag & 46$\pm$7 & 100  & ? \\
$^{130}$Ag & $\approx$50 & ?   & ? \\
$^{130}$Cd & 162$\pm$7  & 100 & 3.5$\pm$1.0 \\
$^{131}$Cd & 68$\pm$3  & 100 & 3.5$\pm$1.0 \\
$^{132}$Cd & 97$\pm$10  & 100 & 60$\pm$15 \\
$^{131}$In & 280$\pm$30   & 100 & $\leq$2.0$\pm$0.3 \\
$^{132}$In & 207$\pm$6   & 100 & 6.3$\pm$0.9 \\
$^{133}$In & 165$\pm$3 & 100 & 85$\pm$10 \\
$^{134}$In & 140$\pm$4 & 100 & 65 \\
$^{135}$In & 92$\pm$10   & 100 & $>$0 \\
$^{130}$Sn & 223200$\pm$4200  & 100 & ? \\
$^{131}$Sn & 56000$\pm$5000  & 100 & ? \\
$^{132}$Sn & 39700$\pm$800  & 100 & ? \\
$^{133}$Sn & 1460$\pm$30  & 100 & 0.0294$\pm$24 \\
$^{134}$Sn & 1050$\pm$11   & 100 & 17$\pm$13 \\
$^{135}$Sn & 530$\pm$20  & 100 & 21$\pm$3 \\
$^{136}$Sn & 250$\pm$30   & 100 & 30$\pm$5 \\
$^{137}$Sn & 190$\pm$60  & 100 & 58$\pm$15 \\
$^{138}$Sn & $>$0.000408 & ?   & ? \\
$^{139}$Sn &  ?  & ?  &  ? \\
$^{140}$Sn &  ?  & ?  & ?  \\
$^{136}$Sb & 923$\pm$14 &  100 & 16.3$\pm$3.2 \\
$^{137}$Sb & 450$\pm$50 &  100 & 49$\pm$10 \\
$^{138}$Sb & $\geq$0.0003  & ? & ? \\
$^{139}$Sb & $>$0.00015   & ? & ?  \\
$^{137}$Te & 2490$\pm$50  & 100 & 2.99$\pm$16 \\
$^{138}$Te & 1400$\pm$400  & 100 & 6.3$\pm$2.1 \\
$^{139}$Te & $>$0.00015   & ? & ? \\
\hline
\end{tabular}
\caption[Properties of neutron-rich nuclides relevant to the A=130 r-process peak.]{Properties of neutron-rich nuclides relevant to the A=130 r-process peak. All data are taken from the ENSDF database ({\it http://www.nndc.bnl.gov/ensdf}) \citep{Burrows90}. The ? symbol was used where data were not available.}\label{Table2}
\end{table}
The regularly updated   ENDF and  ENSDF database evaluations  could provide valuable for  r-process data in the actinide region. The extremely neutron-rich superheavy fission nuclei play an important role in element production. Presently, these nuclei can be studied only with theoretical model calculations. These calculations could be calibrated using the existing actinide data for Maxwellian-averaged neutron cross sections, half-lives, spontaneous fission and delayed neutrons probabilities.

An example of nuclear reaction and structure data sets for the Z=90-110 region is shown in Table \ref{Table3}.  The tabulated values demonstrate an increasingly complex nature of nuclear decay for Z$>$95 nuclei where spontaneous fission and $\beta$-decay play an important role. Spontaneous fission fragments are of interest to the r-process studies. Complimentary information on fission fragments distribution can be obtained from the Sigma Web interface {\it http://www.nndc.bnl.gov/sigma} \citep{Pritychenko08}.
\begin{table}
\centering
\begin{tabular}{|c|ccccc|}
\hline
Isotope &  $\sigma (n,F)$, mb & T$_{1/2}$, y  & SF, $\%$ &   $\alpha$-decay, $\%$ & $\beta$-decay, $\%$\\
\hline \hline
$^{232}$Th & 1.672E1 & 1.40E10 & 1.1E-9 & 100  & ? \\
$^{235}$U &  1.376E3 & 7.04E8 & 7E-9 & 100  & ? \\
$^{238}$U & 7.839E1 & 4.468E9 & 5.45E-5 &  100 & ? \\
$^{237}$Np & 9.953E2 & 2.144E6 & $\leq$2E-10 & 100  & ? \\
$^{239}$Pu & 1.885E3 & 2.411E4 & 3.1E-10 & 100  & ? \\
$^{241}$Am & 7.767E2 & 432.6 & 3.6E-10 & 100 & ? \\
$^{250}$Cm & 3.192E2 & 8.3E3 & 74 & 18  & 8 \\
$^{250}$Bk & 1.209E3 & 3.212 h  & ? & ? &  100  \\
$^{252}$Cf & 2.133E3 & 2.645 & 3.092 & 96.908  & ? \\
$^{255}$Es & 3.108E2 & 39.8 d & 0.0041 & 8 & 92 \\
$^{255}$Fm & 2.776E3 & 20.07 h & 2.4E-5 & 100 & ? \\
\hline
\end{tabular}
\caption[Nuclear reaction and structure properties of several actinides.]{Nuclear reaction and structure properties of several actinides. Reaction cross sections were calculated from ENDF/B-VII.1 library at $kT$=400 keV and decay data were taken from the ENSDF database ({\it http://www.nndc.bnl.gov/ensdf}). The ? symbol was used where data were not available.}\label{Table3}
\end{table}

\subsection{p-process}
A detailed analysis of the Fig. \ref{chart} data indicates between 29 and 35 proton-rich nuclei that cannot be produced in the s- or r-processes. A significant fraction of these nuclei originate from the $\gamma$-process \citep{Woosley78,Boyd08}. This process could take place in Type-II supernovae  at (2-3)$\times$T$_{9}$. It begins with ($\gamma$,n) reactions that synthesize proton-rich heavy nuclei that are followed by charged-particle emitting reactions. Such process includes an extensive reaction network consisting of approximately 20,000 reactions and 2,000 nuclei. Due to lack of experimental data, p-process network calculations are often based on theoretical model predictions. This situation can be improved via addition of the known experimental reaction cross sections.

\begin{figure}[htb]	% h-here, t-top, b-bottom
\centering
\includegraphics[width=12cm]{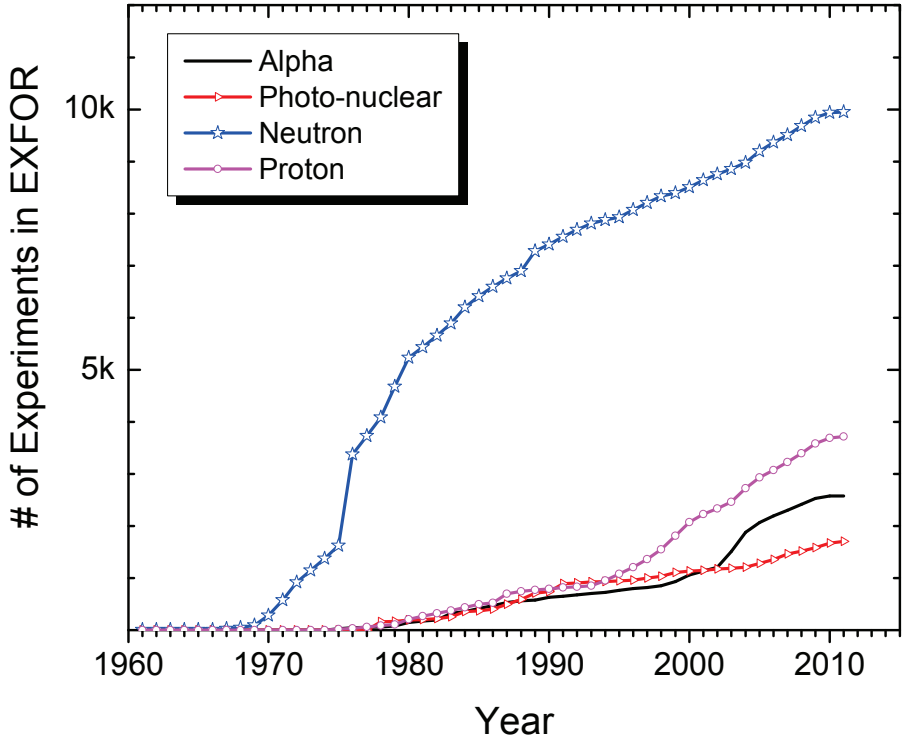} %	** if .eps don't need extension
\caption{Time and content evolution of the EXFOR database. Initially database scope was limited to neutron-induced reaction cross sections, later scope expansion included charge particle and photo-nuclear reactions.}\label{x4}
\end{figure}

These reactions have been compiled  in the EXFOR database since 70ies, and the database content  is shown in Fig. \ref{x4}. For historic reasons, it is relatively complete for neutron-,  proton- and alpha-induced reaction compilations, and has a limited number of compilations for heavy-ion and photonuclear reactions. The IAEA EXFOR Web interface: {\it http://www-nds.iaea.org/exfor}  \citep{Zerkin05} allows user-friendly nuclear astrophysics data search using multiple parameters, such as target, nuclear reaction, cross section, and energy range.
The interface is currently used for a {\it p}-process nuclei data mining operation at ATOMKI   \citep{Szucs10}.  EXFOR is the best source of experimental nuclear reaction data; however, it is not complete, and it takes $\sim$1-2 y before article compilation is completed.

To overcome this problem the Nuclear Science References (NSR) database ({\it http://www.nndc.bnl.gov/nsr}) \citep{Pritychenko11} is recommended.  Table \ref{Table4} shows EXFOR and NSR database content for the ATOMKI project scope of (p,$\gamma$), (p,n), (p,$\alpha$), ($\alpha$,$\gamma$), ($\alpha$,n) and ($\alpha$,p) reactions.
\begin{table}
\centering
\begin{tabular}{|c|cc|}
\hline
Reaction   	  & EXFOR & NSR  \\
\hline \hline
p,$\gamma$ 	  & 396   & 2162 \\
p,n                        & 666   & 2571 \\
p,$\alpha$          & 337   & 1031 \\
$\alpha$,$\gamma$ & 144   & 522  \\
$\alpha$,n        & 343   & 1321 \\
$\alpha$,p        & 166   & 848  \\
\hline
\end{tabular}
\caption[Total number of p-process reaction entries in EXFOR and NSR databases.]{Total number of p-process reaction entries in EXFOR ({\it http://www-nds.iaea.org/exfor}) and NSR ({\it http://www.nndc.bnl.gov/nsr}) databases as of  August 2011.}\label{Table4}
\end{table}
The tabulated data indicate a factor of 3-5 difference between two databases. This is mostly due to the fact that multiple article can be combined into a single EXFOR entry and gaps in the EXFOR coverage.

Another important tool for p-process nucleosynthesis studies is the nuclear reaction reciprocity theorem. It allows extracting a reaction cross section if an inverse reaction is known. For 1 + 2 $\rightarrow$ 3 + 4 and 3 + 4 $\rightarrow$ 1 + 2 processes the cross section ratio is
\begin{equation}
\label{myeq.r6}
\frac{\sigma_{34}}{\sigma_{12}} = \frac{m_3 m_4 E_{34}(2 J_3 + 1) (2 J_4 + 1)(1 + \delta_{12})}{m_1 m_2 E_{12}(2 J_1 + 1) (2 J_1 + 1)(1 + \delta_{34})},
\end{equation}
where $E_{12}$,  $E_{34}$ are kinetic energies in the c.m. system, J is angular momentum, and $\delta_{12}$=$\delta_{34}$=0.

\subsection{rp-process}
The rp-process (rapid proton capture process) consists of consecutive proton captures onto seed nuclei to produce heavier elements. It occurs in a number of astrophysical sites including X-ray bursts, novae, and supernovae. The rp-process  requires   a high-temperature environment ($\sim$10$^{9}$ K) to overcome Coulomb barrier for charged particles. This process contributes to observed abundances  of light and medium proton-rich nuclei and compliments the p-process. The rp-process data needs include
\begin{itemize}
\item nuclear masses
\item proton capture rates
\end{itemize}

\section{Conclusion}
Finally, a review of stellar nucleosynthesis and its data needs has been presented. Several nuclear astrophysics opportunities and the corresponding computation tools and methods have been identified.  Complimentary sets of nuclear data for s-process nucleosynthesis have been produced. These, nuclear astrophysics model-independent data sets are based on the latest evaluated nuclear libraries and low-fidelity covariances data.  This analysis indicates that the nucleosynthesis processes and respective abundances are strongly affected by both nuclear reaction cross sections  and  nuclear structure effects. Several nuclear structure data sets that include $\beta$-decay half-lives, spontaneous fission,  and delayed neutron emission probabilities have been considered.

Present results demonstrate a wide range of uses for nuclear reaction cross sections and structure data in stellar nucleosynthesis. They provide additional benchmarks and build a bridge between nuclear astrophysics and nuclear industry applications. Further work will include extensive data analysis, neutron physics, and network calculations.

We are grateful to M. Herman (BNL)  for the constant support of this project,  to S. Goriely (Universite Libre de Bruxells) and R. Reifarth (Goethe University) for productive discussions, and to V. Unferth (Viterbo University) for a careful reading of the manuscript. This work was sponsored in part by the Office of Nuclear Physics, Office of Science of the U.S. Department of Energy under Contract No. DE-AC02-98CH10886 with Brookhaven Science Associates, LLC.
%\section{References Format}
%References have to include at least 5 items and have not to be self-centred. The list of references has to be arranged alphabetically according to the first author, subsequent lines indented. Do not number references. Publications by the same author(s) should be listed in order of the year of the publication. If there are more than one manuscript by the same author(s) and with the same date, label them a, b, etc. Please note that all references listed here must be directly cited in the body of the text.  \nocite{siegwart01,lima04,li96,arai99}

%%% Without a bib file, write your references like this: **************

%%% With a bib file, include it! *************************************
%\bibliography{temp}

\begin{thebibliography}{100}
\bibitem[Aldama \& Trkov, 2004]{Aldama2004} Aldama, D.L. \& Trkov, A. (2004). FENDL-2.1
Update of an evaluated nuclear data library for fusion applications. \emph{International Atomic Energy Agency}, INDC(NDS)-467, Distribution FE, December 2004.

\bibitem[Anders \& Grevesse, 1989]{Anders89} Anders, E. \& Grevesse, N. (1989). Abundances of the elements - Meteoritic and solar. \emph{Geochimica et Cosmochimica Acta}, Vol. 53, January 1989, pp. 197-214.

\bibitem[Angulo et al., 1999]{Angulo99} Angulo, C.;  Arnould, M.;  Rayet, M.; Descouvemont, P.; Baye, D.; Leclercq-Willain, C.; Coc, A.; Barhoumi, S.; Aguer, P.; Rolfs, C.; Kunz, R.; Hammer, J.W.; Mayer, A.; Paradellis, T.; Kossionides, S.; Chronidou, C.; Spyrou, K.;  Degl'Innocenti, S.; Fiorentini, G.; Ricci, B.; Zavatarelli, S.; Providencia, C.; Wolters, H.; Soares, J.; Grama, C.; Rahighi, J.; Shotter, A. \& Lamehi Rachti, M. (1999). A compilation of charged-particle induced thermonuclear reaction rates. \emph{ Nuclear Physics A}, Vol. 656, No.1, August 1999, pp. 3-183.

\bibitem[Aprahamian, 2011]{Aprahamian11} Aprahamian, A. (2011). R-process mass sensitivities. \emph{Proc. XIV International Symposium on Capture $\gamma$-ray Spectroscopy}, August 28-September 2, 2011, University of Guelph, Guelph, Ontario, Canada.


\bibitem[Arlandini et al., 1999]{Arlandini99}  Arlandini, C.; K\"{a}ppeler, F. ; Wisshak, K.; Gallino, R.; Lugaro, M.; Busso, M. \& Straniero, O. (1999). Neutron Capture in Low-Mass Asymptotic Giant Branch Stars: Cross Sections and Abundance Signatures.  \emph{The Astrophysical  Journal},
Vol. 525, November 1999, pp. 886-900.

\bibitem[Audi et al., 2003]{Audi03} Audi, G.; Wapstra, A.H. \& Thibault, C. (2003). The AME2003 atomic mass Evaluation (II). Tables, graphs, and references. \emph{Nuclear Physics A}, Vol. 729, Issue 1, December 2003, pp. 337-676.

\bibitem[Audi \& Meng, 2011]{Audi11}  Audi, G. \& Meng, W. (2011). Atomic Mass Evaluation 2011. \emph{Private Communication}.

\bibitem[Bao et al., 2000]{Bao00} Bao, Z.Y.; Beer, H.; K\"{a}ppeler, F.; Voss, F.;  Wisshak, K.; Rauscher, T. (2000). Neutron Cross Sections for Nucleosynthesis Studies. \emph{Atomic Data and Nuclear Data Tables}, Vol. 76, Issue 1, September 2000, pp. 70-154.

\bibitem[Bethe, 1939]{Bethe39} Bethe, H.A. (1939). Energy Production in Stars. \emph{Physical Review}, Vol. 55, Issue5, March 1939, pp. 434-456.

\bibitem[Boyd, 2008]{Boyd08} Boyd, R.N.  (2008). \emph{An introduction to nuclear astrophysics}, The University of Chicago Press, 2008.

\bibitem[Boyd et al., 2009]{Boyd09} Boyd, R.N; Bernstein, L. \& Brune, C. (2009). Studying Nuclear Astrophysics at NIF. \emph{Physics Today}, August 14, 2009.

\bibitem[Burbidge et al., 1957]{Burbidge57} Burbidge, E.M.; Burbidge, G.R.; Fowler, W.A. \& Hoyle, F. (1957). Synthesis of the Elements in Stars. \emph{Review of Modern Physics}, Vol. 29, No.4, October 1957, pp. 547-654.

\bibitem[Burrows, 1990]{Burrows90} Burrows, T.W. (1990). The evaluated nuclear structure data file: Philosophy, content, and uses. \emph{Nuclear Instruments and Methods in Physics Research Section A}, Vol. 286, Issue 3, January 1990, pp. 595-600.

\bibitem[Cameron, 1957]{Cameron57} Cameron, A.G.W. (1957). Stellar evolution, nuclear astrophysics, and nucleogenesis. \emph{Chalk River Report}, CLR-41, Chalk River, Ontario, June 1957, pp. 1-197.

\bibitem[Chadwick et al., 2006]{Chadwick06}  Chadwick, M.B.; Oblo{\v z}insk{\' y}, P.; Herman, M.; Greene, N.M.; McKnight, R.D.;  Smith, D.L.; Young, P.G.; MacFarlane, R.E.; Hale, G.M.; Frankle, S.C.; Kahler, A.C.; Kawano, T.; Little, R.C.; Madland, D.G.; Moller, P.; Mosteller, R.D.; Page, P.R.; Talou, P.; Trellue, H.; White, M.C.; Wilson, W.B.; Arcilla, R.; Dunford, C.L.; Mughabghab, S.F.; Pritychenko, B.; Rochman, D.; Sonzogni, A.A.; Lubitz, C.R.; Trumbull, T.H.; Weinman, J.P.; Brown, D.A.; Cullen, D.E., Heinrichs, D.P.; McNabb, D.P.; Derrien, H.; Dunn, M.E.; Larson, N.M.; Leal, L.C.; Carlson, A.D.; Block, R.C.; Briggs, J.B.; Cheng, E.T.; Huria, H.C.; Zerkle, M.L.; Kozier, K.S.; Courcelle, A.; Pronyaev, V. \& van der Marck, S.C. (2006). ENDF/B-VII.0: Next Generation Evaluated Nuclear Data Library for Nuclear Science and Technology. \emph{Nuclear Data Sheets},  Vol. 107, Issue 12, December 2006, pp. 2931-3060.


\bibitem[Cowan \& Thielemann, 2004]{Cowan04} Cowan, J. J. \& Thielemann, F.-K. (2004) R-process Nucleosynthesis in Supernovae. \emph{Physics Today} Vol. 57, October 2004, pp. 47-53.

\bibitem[Cyburt et al., 2010]{Cyburt10} Cyburt, R.H.; Amthor, A.M.; Ferguson, R.; Meisel, Z.; Smith, K.; Warren, S.; Heger, A.; Hoffman, R.D., Rauscher, T.; Sakharuk, A.; Schatz, H.; Thielemann, F.L. \& Wiescher, M. (2010). The JINA REACLIB Database: Its Recent Updates and Impact on Type-I X-ray Bursts. \emph{The Astrophysical Journal Supplement Series}, Vol. 189, July 2010, pp. 240-252.

\bibitem[Dillmann et al., 2006]{Dillmann06}  Dillmann, I.; Heil, M.;  K\"{a}ppeler, F.;  Plag, R.;  Rauscher , T. \&  Thielemann, F.-K. (2006). KADoNiS - The Karlsruhe Astrophysical Database of Nucleosynthesis in Stars. \emph{Proceedings of Capture Gamma-ray Spectroscopy and Related Topics: 12$^{th}$ International Symposium}, Vol. 819, pp. 123-127, ISBN: 0-7354-0313-9,  Notre Dame, Indiana (USA), 4-9 September 2005, AIP Conference Proceedings.

\bibitem[Eddington, 1920]{Eddington20} Eddington, A.S. (1920). The Internal Constitution of the Stars. \emph{Nature}, Vol. 106, September 1920, pp. 14-20.

\bibitem[Grevesse \&  Sauval, 1998]{Grevesse98} Grevesse, N. \& Sauval, A. J. (1998). 	 Standard Solar Composition. \emph{Space Science Reviews}, v. 85, Issue 1/2, 1998, pp. 161-174.

\bibitem[Herman et al., 2007]{Herman07} Herman, M.; Capote, R.; Carlson, B.V.;  Oblo{\v z}insk{\' y}, P.; Sin, M.;  Trkov, A.;  Wienke, H. \& Zerkin, V. (2007). EMPIRE: Nuclear Reaction Model Code System for Data Evaluation.  \emph{Nuclear Data Sheets},  Vol. 108, Issue 12, December 2007, pp. 2655-2715.

\bibitem[Hoyle, 1946] {Hoyle46} Hoyle, F. (1946). The synthesis of the elements from hydrogen. \emph{Monthly  Notices of the Royal Astronomical  Society} Vol. 106, 1946, pp. 343-383.


\bibitem[K\"{a}ppeler,  2011]{FKaeppeler11}  K\"{a}ppeler, F. (2011). Reaction cross sections for s, r, and p process. \emph{Progress in Particle and Nuclear Physics}, Vol. 66, Issue 2, April 2011, pp. 390-399.

\bibitem[K\"{a}ppeler et al., 2011]{Kaeppeler11}  K\"{a}ppeler, F.; Gallino, R.; Bisterzo, S. \& Wako Aoki. (2011). The s-process: Nuclear physics, stellar models, and observations.  \emph{Review of Modern Physics}, Vol. 83, January-March 2011, pp. 157-193.

\bibitem[Kolb \& Turner, 1988]{Kolb88} Kolb, E. \& Turner, M. (1988). \emph{The Early Universe}, Addison-Wesley, 1988.

\bibitem[Koning et al., 2008]{Koning08} Koning, A.J.; Hilaire, S, \& Duijvestijn, M.C. (2008). TALYS-1.0.  \emph{Proceedings of International Conference on Nuclear Data for Science and Technology}, Vol. 1, pp. 211-214, ISBN:978-2-7598-0090-2, Nice, France, April 22-27, 2007, EDP Sciences, 2008.

 \bibitem[Little et al., 2008]{Little08} Little, R.C.; Kawano, T.; Hale, G.D.; Pigni, M.T.; Herman, M.; Oblozinsky, P.; Williams, M.L.; Dunn, M.E.; Arbanas, G.; Wiarda, D.; McKnight, R.D.; McKamy, J.N. \& Felty, J.R. (2008). Low-fidelity Covariance Project.  \emph{Nuclear Data Sheets}, Vol. 109, Issue 12, December 2008, pp. 2828-2833.

 \bibitem[Merrill, 1952]{Merrill52}  Merrill, S. P. W. (1952). Spectroscopic Observations of Stars of Class S. \emph{The Astrophysical Journal}, Vol. 116, February 1952, pp. 21-26.


  \bibitem[Mughabghab, 2006]{Mughabghab06} Mughabghab, S.F. (2006). \emph{Atlas of Neutron Resonances, Resonance Parameters and Thermal Cross Sections. Z=1-100}, Elsevier, 2006.

\bibitem[Nakagawa et al., 2005]{Nakagawa05} Nakagawa, T.;  Chiba, S; Hayakawa, T. \& Kajino, T. (2005).  Maxwellian-averaged neutron-induced reaction cross sections and astrophysical reaction rates for kT = 1 keV to 1 MeV calculated from microscopic neutron cross section library JENDL-3.3.  \emph{Atomic Data and Nuclear Data Tables},  Vol. 91, Issue 2, November 2005, pp. 77-186.

\bibitem[NRDC, 2011]{NRDC11} NRDC, International Network of Nuclear Reaction Data Centres. (2011). \emph{Compilation of experimental nuclear reaction data	(EXFOR/CSISRS)}, (Available	from	{\it http://www- nds.iaea.org/exfor/}, {\it http://www.nndc.bnl.gov/exfor/}).

\bibitem[Pritychenko et al., 2006]{Pritychenko06} Pritychenko, B.;  Sonzogni, A.A.; Winchell, D.F.; Zerkin, V.V; Arcilla, R.; Burrows, T.W.; Dunford, C.L.; Herman, M.W.; McLane, V.; Oblo{\v z}insk{\' y}, P.; Sanborn, Y. \& Tuli, J.K. (2006). Nuclear Reaction and Structure Data Services of the National Nuclear Data Center. \emph{Annals of Nuclear Energy}, Vol. 33, Issue 4, March 2006, pp. 390-399.

\bibitem[Pritychenko \& Sonzogni, 2008]{Pritychenko08} Pritychenko, B \& Sonzogni, A.A. (2008). Sigma: Web Retrieval Interface for Nuclear Reaction Data. \emph{Nuclear Data Sheets}, Vol. 109, Issue 12, December 2008,  pp. 2822 - 2827.

\bibitem[Pritychenko et al., 2010]{Pritychenko10}  Pritychenko, B.;  Mughabghab, S.F. \&  Sonzogni, A.A. (2010). Calculations of Maxwellian-averaged Cross Sections and Astrophysical Reaction Rates Using the ENDF/B-VII.0, JEFF-3.1, JENDL-3.3 and ENDF/B-VI.8 Evaluated Nuclear Reaction Data Libraries. \emph{Atomic Data and Nuclear Data Tables}, Vol. 96, Issue 6, November 2010, pp. 645-748.

\bibitem[Pritychenko, 2010]{BPritychenko10}  Pritychenko, B. (2010). Complete calculation of evaluated Maxwellian-averaged cross sections and their uncertainties for s-process nucleosynthesis. \emph{Proceedings of  11th Symposium on Nuclei in the Cosmos, NIC XI}, PoS(NIC-XI)197, Heidelberg, Germany, July 19-23, 2010, Proceedings of Science, 2010.

\bibitem[Pritychenko et al., 2011]{Pritychenko11}  Pritychenko, B.; B{\v e}t{\' a}k, E.; Kellett, M.A.; Singh, B. \& Totans, J. (2011). The Nuclear Science References (NSR) database and Web Retrieval System. \emph{Nuclear Instruments and Methods in Physics Research Section A}, Vol. 640, Issue 1, 1 June 2011, pp. 213-218.

\bibitem[Rauscher \& Thielemann, 2000]{Rauscher00} Rauscher T. \& Thielemann F. (2000). Astrophysical Reaction Rates from Statistical Model Calculations. \emph{Atomic Data and Nuclear Data Tables}, Vol.75, Issue 1, January 2000, 1.

\bibitem[Rolfs \& Rodney, 1988]{Rolfs88} Rolfs, C.E. \& Rodney, W.S. (1988). \emph{Cauldrons in the Cosmos}, The University of Chicago Press, 1988.


\bibitem[Shibata et al., 2011]{Shibata01} Shibata, K.;  Iwamoto, O.; Nakagawa, T.; Iwamoto, N; Ichihara, A.; Kunieda, S.; Chiba, S.; Furutaka, K.; Otuka, N.; Ohsawa, T.; Murata, T.; Matsunobu, H.; Zukeran, A.; Kamada, S. \& Katakura, J.-i. (2011). JENDL-4.0: A New Library for Nuclear Science and Engineering. \emph{Journal of Nuclear Science and Technology}, Vol. 48, No. 1, January 2011, pp. 1-30.

\bibitem[Sz\"{u}cs et al., 2010]{Szucs10} Sz\"{u}cs, T.; Dillmann, I.;  Plag, R. \& F\"{u}l\"{o}p, Zs. (2010).  The new p-process database of KADoNiS. \emph{Proceedings of  11$^{th}$ Symposium on Nuclei in the Cosmos, NIC XI} PoS(NIC-XI)247, Heidelberg, Germany, July 19-23, 2010, Proceedings of Science, 2010.

\bibitem[Wolram, 2011]{Wolf11} Wolfram Corporation. (2011). \emph{Wolfram Mathematica Online Integrator}, Available from  $\langle$http://integrals.wolfram.com$\rangle$.

\bibitem[Woosley \&  Howard, 1978]{Woosley78}  Woosley, S. E.  \& Howard, W. M. (1978). The p-process in supernovae. \emph{The Astrophysical Journal Supplement}, Vol. 36, 1978, pp. 285-304.

\bibitem[Zabrodskaya et al., 2007]{Zabrodskaya07} Zabrodskaya, S.V.; Ignatyuk, A.V.; Koscheev, V.N.; Manokhin, V.N.; Nikolaev, M.N. \& Pronyaev, V.G. (2007). ROSFOND - Russian National Library of Neutron Data. \emph{Voprosy Atomnoj Nauki i Techniki, Seriya: Nuclear Constants} Vol. 1-2, 2007, p. 3-21 .

\bibitem[Zerkin et al., 2005]{Zerkin05}  Zerkin, V.V.;  McLane, V.;  Herman, M.W. \& Dunford, C.L. (2005). EXFOR-CINDA-ENDF: Migration of Databases to Give Higher-Quality Nuclear Data Services. \emph{Proceedings of International Conference on Nuclear Data for Science and Technology}, Vol. 769, pp. 586-589, ISBN: 0-7354-0254-X, Santa Fe, New Mexico (USA), 26 September-1 October 2004, AIP Conference Proceedings.





\end{thebibliography}

\end{document}